\documentclass[runningheads]{llncs}

\usepackage{graphicx}
\usepackage{todonotes}
\usepackage{enumitem}
\usepackage{amsmath}

\usepackage{pgfplots}
\usepackage{adjustbox}
\usepackage{listings,multicol, multirow}
\usepackage{xcolor,colortbl}

\usepackage{array}

\pgfplotsset{compat=1.16}
\usepgfplotslibrary{statistics}

\begin{document}

\title{Bug Prediction Using Source Code Embedding Based on Doc2Vec}

\author{
Tamás Aladics\inst{1}\orcidID{0000-0002-4689-8878} \and
Judit Jász\inst{1,2}\orcidID{0000-0001-6176-9401} \and
Rudolf Ferenc\inst{1}\orcidID{0000-0001-8897-7403}
}

\authorrunning{T.Aladics et al.}

\institute{University of Szeged, Hungary \\
	\and FrontEndART Ltd., Hungary \\
\email{\{aladics,jasy,ferenc\}@inf.u-szeged.hu}}
\maketitle

\abstract{Bug prediction is a resource demanding task that is hard to automate using static source code analysis. In many fields of computer science, machine learning has proven to be extremely useful in tasks like this, however, for it to work we need a way to use source code as input. We propose a simple, but meaningful representation for source code based on its abstract syntax tree and the Doc2Vec embedding algorithm. This representation maps the source code to a fixed length vector which can be used for various upstream tasks -- one of which is bug prediction. We measured this approach's validity by itself and its effectiveness compared to bug prediction based solely on code metrics. We also experimented on numerous machine learning approaches to check the connection between different embedding parameters with different machine learning models. Our results show that this representation provides meaningful information as it improves the bug prediction accuracy in most cases, and is always at least as good as only using code metrics as features.}


\section{\uppercase{Introduction}}
\label{sec:introduction}

Detecting bugs is one of the most important steps in software development to ensure the quality of the product. It is a challenging task that varies greatly in difficulty based on the code, and it is very resource demanding.  Numerous tools are used in the field of static code analysis to leverage this problem, but most of these lack robustness as they are mostly used to find manually defined patterns. Machine learning algorithms can help in improving the robustness of bug prediction by learning patterns from big quantities of examples. 

One of the fundamental aspects of any machine learning method is the way the input features are generated. In the case of software related tasks, these features are mostly derived from the source code: code metrics, token sequences of functions, classes, or whole programs. There are numerous ways of source code representation, which we briefly introduce in the next section. These approaches vary in the structure chosen for the representation (token, statement, function, class, etc.), and in the form the code is used (as text, abstract syntax tree, etc.). 

However, most of the related code representation methods are not used in bug prediction tasks. As shown in the related literature, it is apparent that code metrics can be used for bug prediction tasks and achieve relatively good results~\cite{FBG20,HHA18,PDC16}.
In this work we are looking for the answer to how different results we can get during bug prediction by replacing traditional software metrics with feature vectors extracted from the abstract syntax tree (AST) in the learning process.

More precisely, we represent source code by traversing the AST in depth-first manner, and thus generate a sequence of tokens, on which we train numerous Doc2Vec models with different parameters. We then use these trained models to generate fixed length vectors for each Java class's source code. This vector can be used as a standalone feature vector or it can be supplemented with metrics. We discuss the details of our methodology in Section~\ref{sec:methodology}.

Finally, we validated the effectiveness of this source code representation to predict bugs on the Unified Bug Dataset~\cite{FTL20}, which is a dataset of buggy and non-buggy classes implemented in Java. We report our results in Section~\ref{sec:results}, where we also sought answers to our research questions regarding this representation:

\begin{description}
 \item[RQ1] Is there a Doc2Vec parametrization that would produce similar or better results than learning based on code metrics?
 \item[RQ2] Can we improve performance by combining source code embedding and code metrics?
 \item[RQ3] Do the features provided by source code embedding give valid information or was it only random noise learned?
\end{description}

\noindent
Our main contributions are the following:
\begin{itemize}
\item A simple and scalable method to embed source code into a fixed length vector.
\item An empirical evaluation of source code embedding with various machine learning models.
\item A detailed discussion on the effects of using source code embedding with and without code metrics.
\end{itemize}

\section{\uppercase{Related Work}}
\label{sec:related}

In the field of source code analysis, different methods have been proposed to find ways to represent source code. One way to categorize these approaches is granularity, namely, what is the basic structure that is chosen to generate this representation - typically tokens, functions or classes. This categorization is applied by~\cite{chen2019literature} which we present in a brief overview of representation methods, focusing on the function and token embeddings, as our approach is a mix of the two. 

The most fundamental structures that can be used as a basis for representation are tokens. \cite{cpp_tokenbased} uses tokens as inputs for different kinds of models, one of which is Word2Vec, to generate word embedding for C/C++ tokens for software vulnerability prediction. Stepping one step up in granularity, \cite{python_functionbased} use function embedding as input to repair variable misuse in Python. They encode the function AST by doing a depth first traversal of its nodes and create an embedding by concatenating the absolute position of the node, the type of node, the relationship between the node and its parent, and the string label of the node. In another work, ~\cite{c_functionbased} generate function embeddings for C code using control-flow graphs. They perform a random walk on interprocedural paths in the program, and use the paths to generate function embeddings. \cite{CMB19} used a specific CNN (Convolutional Neural Network) architecture to extract features from the AST, and then predicted bugs with logistic regression. There are works that use structures of wider scope like compilation units, but since our representation does not use these, we do not discuss them here.

Extracting features based on the source code directly is a good way to gather local information about the chosen piece of code. However, using code metrics can provide an additional, more global description of the code and its environment. Significant research has happened in this direction as well, where typically a number of metrics are derived from source code or based on results generated by test suites.

Concretely,~\cite{FBG20} use metrics such as LOC (Lines of Code), TCD (Total Comment Density), and NL (Nesting Level) etc. as input to a set of machine learning models, and they also provided comparisons for them based on the results. \cite{HHA18} used Naive Bayes, Neural Networks, and Decision Trees on datasets generated by testing processes. \cite{PDC16} predicted bugs on the Eclipse JDT, using a set of 4 metrics.

Researchers also demonstrated the usefulness of using AST in software engineering tasks such as code completion.
\cite{WLT16} leveraged Deep Belief Network (DBN) to automatically learn semantic features from token vectors extracted from programs' ASTs. \cite{SBH19} use AST n-grams to identify features of defective Java code that improve defect prediction performance.
 
As already mentioned, related research shows that using natural language processing (NLP) methods, such as Word2Vec on source code can be promising. The now widespread Word2Vec method is a way to generate meaningful vector representations for words in text proposed by~\cite{mikolov2013efficient}. Doc2Vec is a natural extension to Word2Vec published by \cite{mikolov2013distributed}, which introduces distributed representation for documents, that is, sets of words.

Our proposed method is using a mix of the previously introduced ideas, which we explain more thoroughly in the next section.

\section{\uppercase{Methodology}}
\label{sec:methodology}

Our main objective in this paper is to create a code embedding based bug prediction model, and compare it with existing bug prediction methods based on code metrics.
The brief overview of our method is as follows: we use class-like elements of Java codes (which means classes, enumerations, and interfaces, to which we refer to as  classes in the future) as a basic structure to generate sequences. Then, we treat each of these sequences (corresponding to classes) as documents in a Doc2Vec model, and the tokens making up the classes as words making up the documents. 
Doc2vec assigns to each class a fixed size vector as output, which are thus suitable for inputs of the learning task.
In our experiments we generate document vectors with different parametrizations, on which we perform bug prediction, using them as standalone features, or supplement them with code metrics.

\subsection{AST embedding}

When considering source code representations, the first question is the form of the source code we work with. 
We decided to take an intermediate representation used by compilers, the abstract syntax tree (AST), because it captures structural information, which is an important aspect of programs.
However, an AST is a tree structure, and we need a numeric representation for it.

\label{sec:ast_embedding}

\vspace{-0.5cm}\begin{figure}[th]
	\centering
	\small\begin{lstlisting}
       public class Debug {
         public static boolean isDebugOn=true;
         public static void debug(String s) {
           if (isDebugOn) {
             System.out.println(s);
           }
         }
       }

	\end{lstlisting}
	\caption{Example Java code}
	\label{fig:CG_java_example}
\end{figure} 

\begin{figure*}[thb!]
	\centering
	\includegraphics[width=0.85\textwidth]{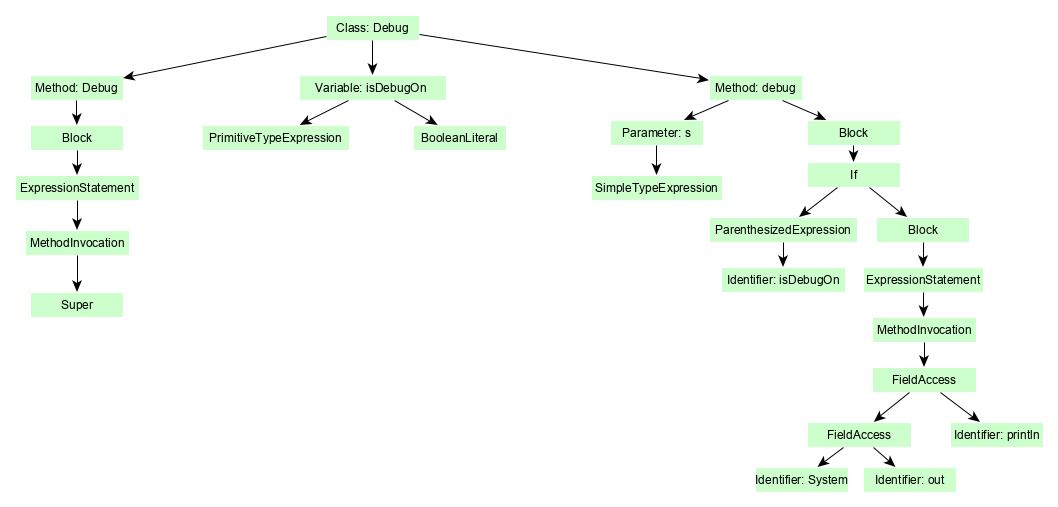}
	\caption{The AST representation of the example Java code}
	\label{fig:CG_java_example_Ast}
\end{figure*}

One simple way to do this is to traverse the tree in a depth-first manner, and add each node to the sequence. Another question is how to map the nodes, since they contain specific data regarding the corresponding code part. For example, a node can be an element with the name ``x'', integer variable etc. In our representation -- to remain general, and to limit the vocabulary size -- we mapped each node to its kind (e.g. class definition, variable usage, assignment operator).

A problem with simple depth-first traversal is that it does not reflect scope changes. That is, for example, if we take an \texttt{if} statement with an expression following:
\begin{lstlisting}
             if(cond){
               expr1;
             }
             expr2;
\end{lstlisting}
\noindent and if we take an \texttt{if} statement with two expressions in its body:
\begin{lstlisting}
             if(cond){
               expr1;
               expr2;
             }	
\end{lstlisting}

\noindent they would map to the same sequences. To take this into account, we added a constant value to the sequence each time a step back occurs in the tree, which is approximately equivalent to scope changes. 
This way we can generate variable length vectors from functions while keeping structural information.
 
Figures~\ref{fig:CG_java_example} and~\ref{fig:CG_java_example_Ast} introduce a very simple Java class and its AST representation.
The nodes in the AST are the source code elements of the example program. 

In our implementation each node kind is represented by a positive number, while the constant value that denotes scope change in the AST could be any non-positive number (to avoid conflicts with node kinds), so we chose -2.
So the example class is represented as a sequence of numbers: 85, 112, 42, -2, 15, -2, -2, 100, 42, -2, 107, 44, -2, -2, 57, 68, 39, 26, -2, -2, 57, 65, 33, 24, 24, 26, -2, 26, -2, -2, 26, -2, -2, 26, -2, -2, -2, -2, -2, -2, -2, -2.

\subsection{Doc2Vec}
\label{sec:doc2vec}

Word2Vec~\cite{mikolov2013efficient} has seen great success as a word embedding algorithm, and Doc2Vec~\cite{mikolov2013distributed} is a natural extension to it for documents. While Word2Vec learns global representations for the words of the vocabulary, Doc2Vec also tries to learn a representation vector for a set of words, which is referred to as paragraph or document. This is done by adding a designated vector (paragraph vector) to be learned along with the set of words making up the paragraph. 

Learning the word vectors and the paragraph vectors can be done in two ways: by using the Distributed Memory version of Paragraph Vector (PV-DM) or the Distributed Bag of Words version of Paragraph Vector (DBOW). 

PV-DM is built upon Word2Vec's continuous bag-of-words (CBOW) implementation with the extension of using document vectors. It learns by trying to predict the missing word (target) based on a set of words (context) and the paragraph vector, which represents the document that the context was sampled from. To phrase it differently, word vectors represent the concept of a word while the document vector represents the concept of a document.

The DV-DBOW algorithm is the Doc2Vec extension of the other approach to train Word2Vec embedding: Skip-Gram. It uses only the paragraph vector as input, and learns to predict a randomly sampled set of words from the paragraph, so it does not learn global representations for the words, only for the paragraphs. In DV-DBOW the word vectors are not learned nor stored and according to the authors' findings in \cite{mikolov2013distributed}, DV-DBOW is computationally better and faster than PV-DM, at the same time it tends to perform a little worse in general than PV-DM, even though for some tasks PV-DBOW can be better. 

We chose to use Doc2Vec because we found that it is very reasonable to think of classes as ``paragraphs'' and source code elements as ``words'' that make up the source code itself. By learning to represent classes based on the code elements making it up, we hoped to get a representation that preserves some of the semantic connection between those elements. Another consideration, which is more technical, is that Doc2Vec outputs fixed length vectors, which prevents the need of taking care of the complications when dealing with variable length inputs for machine learning models.

As most NLP models, Doc2Vec is trained on a corpus too. In our case, we chose to use all of the source code that we had for the bug prediction task coming from the Unified Bug Dataset.

\section{\uppercase{Experiments}}
\label{sec:experiments}

Our experiments are based on the work of~\cite{FBG20}. 
We partially rely on their findings, in terms of which learning parameters yield the best results for a given code metric. 
At the same time, we supplement their results by comparing and combining them with the approach we present.
So we used the database they had used for our experiments. 
These experiments were evaluated by the Deep Water Framework (DWF)~\cite{FVA20}, which includes the Deep Bug Hunter framework developed by them.

For reproducibility of the results presented in the following subsections, we summarize the most important features of the datasets used and how we used DWF for the evaluation.

\subsection{Datasets}
In order to be able to predict errors in software with different ML techniques, we need a dataset of the right size and quality.
The Unified Bug Dataset~\cite{FTL20} is suitable for this purpose. 
This dataset merges several datasets, which are the GitHub Bug Dataset~\cite{TGF16}, the Promise~\cite{JuM10} dataset, and the Bug Prediction Dataset~\cite{DLR10}.
In this unified dataset, bugs found in Java code are assigned to source code elements at different levels, such as file, class, and method levels.

In our experiment, we examine all errors assigned to classes, interfaces, and enumerations, however, we do not treat nested classes as separate entities. 
As a result, not all of our results will be exactly what~\cite{FBG20} got, as they worked differently at these points.
In our case, 48719 classes (including interfaces and enumerations) were analyzed, of which 8242 were faulty.

\subsection{Code metrics}
In our experiments, we did not want to analyze which software metrics are most suitable for bug prediction, so we used the same metrics what were used by~\cite{FBG20} in their experiments.
Even though the Unified Bug Dataset contains the metrics for it's entries, we wanted to have the most up to date versions of them. 
For this reason we incorporated OpenStaticAnalyzer toolset~\cite{osa}, as they did.
The used metrics are listed in Table~\ref{tab_metrics}.

\begin{table}[htbp]
	\centering
	\caption{Metrics calculated by the OpenStaticAnalyzer}
	\resizebox{0.65\textwidth}{!}{
		\begin{tabular}{llrll}
			\textbf{Abbr.} & \textbf{Name} &       & \textbf{Abbr.} & \textbf{Name} \\
			AD    & API Documentation &       & NOA   & Number of Ancestors \\
			CBO   & Coupling Between Object classes &       & NOC   & Number of Children \\
			CBOI  & Coupling Between Object classes Inverse &       & NOD   & Number of Descendants \\
			CC    & Clone Coverage &       & NOI   & Number of Outgoing Invocations \\
			CCL   & Clone Classes &       & NOP   & Number of Parents \\
			CCO   & Clone Complexity &       & NOS   & Number of Statements \\
			CD    & Comment Density &       & NPA   & Number of Public Attributes \\
			CI    & Clone Instances &       & NPM   & Number of Public Methods \\
			CLC   & Clone Line Coverage &       & NS    & Number of Setters \\
			CLLC  & Clone Logical Line Coverage &       & PDA   & Public Documented API \\
			CLOC  & Comment Lines of Code &       & PUA   & Public Undocumented API \\
			DIT   & Depth of Inheritance Tree &       & RFC   & Response set For Class \\
			DLOC  & Documentation Lines of Code &       & TCD   & Total Comment Density \\
			LCOM5 & Lack of Cohesion in Methods 5 &       & TCLOC & Total Comment Lines of Code \\
			LDC   & Lines of Duplicated Code &       & TLLOC & Total Logical Lines of Code \\
			LLDC  & Logical Lines of Duplicated Code &       & TLOC  & Total Lines of Code \\
			LLOC  & Logical Lines of Code &       & TNA   & Total Number of Attributes \\
			LOC   & Lines of Code &       & TNG   & Total Number of Getters \\
			NA    & Number of Attributes &       & TNLA  & Total Number of Local Attributes \\
			NG    & Number of Getters &       & TNLG  & Total Number of Local Getters \\
			NII   & Number of Incoming Invocations &       & TNLM  & Total Number of Local Methods \\
			NL    & Nesting Level &       & TNLPA & Total Number of Local Public Attributes \\
			NLA   & Number of Local Attributes &       & TNLPM & Total Number of Local Public Methods \\
			NLE   & Nesting Level Else-If &       & TNLS  & Total Number of Local Setters \\
			NLG   & Number of Local Getters &       & TNM   & Total Number of Methods \\
			NLM   & Number of Local Methods &       & TNOS  & Total Number of Statements \\
			NLPA  & Number of Local Public Attributes &       & TNPA  & Total Number of Public Attributes \\
			NLPM  & Number of Local Public Methods &       & TNPM  & Total Number of Public Methods \\
			NLS   & Number of Local Setters &       & TNS   & Total Number of Setters \\
			NM    & Number of Methods &       & WMC   & Weighted Methods per Class \\
	\end{tabular}}%
	\label{tab_metrics}%
\end{table}%

\subsection{Deep Water Framework}
To speed up the extensive hyper parameter searching process in our experiments, we used the Deep Water Framework by~\cite{FVA20}, which supports defining arbitrary feature extraction and learning methods for an input dataset, and helps in executing all the training tasks in a distributed manner. It also provides a simple overview of results, which can be used to compare the different feature extraction and learning model combinations. In our exact case, the feature extraction part was generating the input by using Doc2Vec on the sequences from the AST (to which process we refer as AST flattening) and/or metrics, while the learning part consists of the numerous ML models described in the following part of this section.

In the Deep Water Framework the F-score is calculated by using the already implemented metric scoring methods in Sklearn and Tensorflow (in case of the neural network based approaches). The performance evaluation was done by using 10 fold cross validation, where the folds were generated by Sklearn's StratifiedKFold in the model\_selection module.

To achieve different embeddings, we tried a number of Doc2Vec paramet\-ri\-za\-tions beforehand to have an idea about the general setup we should use in our experiments. We have also taken into account the experiences reported by the NLP community using Doc2Vec regarding vector and window sizes. Based on these, the Doc2Vec models were para\-metri\-zed the following ways:
\begin{itemize}
	\item \textbf{method}: DBOW or DV-DM
	\item \textbf{vector size}: 25, 50, 75, 150
	\item \textbf{window size}: 4, 8, 12
	\item \textbf{epoch}: 6, 10, 20, 40, 60, 80, 100
\end{itemize}

We produced different vector representations of the flattened source codes with each combination of these parameters, and we looked at how each learning algorithm works on these.

In the case of the learning algorithms, we did not deal with their different parametrizations, as our goal was to find the most appropriate code embedding, so in the case of the learners we used the settings provided by~\cite{FBG20}.
We experimented with some traditional machine learning methods. Their brief description and their parameters are follows:
\begin{description}
	\item[Random Forest] Random forest is an ensemble method that has seen big success in the realm of traditional (ie. non deep learner) machine learning algorithms. We performed our tests with the following parameters: max depth for each tree: 10, splitting criterion: entropy, number of trees: 100
	\item[Decision Tree] Decision tree is a  machine learning approach with tremendous research background behind it. It is a simpler model, and is typically used in ensembles. For our tests we used decision trees with a max depth of 10 with the gini splitting criterion.
	\item[KNN] K-Nearest Neighbour is a method that classifies entries based on the classes of their K closest neighbors. It is a method that typically scales badly, however, our dataset is rather small, and we can choose the embedding dimension. We classified points with a K value of 18, and uniform distance weights.
	\item[SVM] A support vector machine is a supervised machine learning model that uses classification algorithms for two-group classification problems.  We chose the Radial Basis Function as kernel function, the gamma value was 0.02, while C was 2.6 as penalty value.
	\item[Naive Bayes] A simple method based on the Bayes theorem. It is rather simple without any hyper parameters, and can be viewed as a baseline for our experiments. 
	\item[Linear] Linear Classifier, another baseline method that fits a linear model with coefficients to minimize the residual sum of squares between the observed targets in the dataset, and the targets predicted by the linear approximation. Usually it is used for regression, here it is binned as above or below 0.5 to make a classifier.  
	\item[Logistic] The well studied logistic regression method, which is widely used for classification. We used it with L2 penalty, liblinear solver, regularization weight (C) of 2.0, and with 0.0001 tolerance value for stopping criteria.
\end{description}

And we also investigated two simple neural network architectures:
\begin{description}
	\item[SDNNC] Standard Deep Neural Network. A feed forward network with 5 layers, 200 neurons per layer, and an initial learning rate of 0.05. For all of the intermediate layers ReLu activation, for the output layer sigmoid function was used because the task is binary classification. The loss is accordingly binary crossentropy, the optimizer is AdaGrad. The training was done in 10 epochs with a batch size of 100. 
	\item[CDNNC] Custom Deep Neural Network. A feed forward net with similar configuration as the SDNNC - 5 layers, 250 neurons per layer, and L2 regularization is applied with 0.0005 beta value. During training F-score based early stopping is used. 
\end{description}

Tests with both groups of machine learning approaches were performed using DWF, which uses \emph{scikit-learn} \cite{scikit-learn} as backbone for the traditional, and \emph{Tensorflow}~\cite{tensorflow} for the neural network based approaches.

Due to the peculiarities of the data set, we also applied the preprocessing strategy of~\cite{FBG20} in all cases. 
Thus, in addition to binarization and standardization of the data, 50\% upsampling was also utilized on the training data, using Sklearn's resample class. 

With the DWF we could easily define these tasks while continuously checking the progression. We could set the preprocessing we discussed to all of the experiments, then we defined a set of feature extraction methods and another set of machine learning models. DWF then executed all of them combined pairwise. 
The actual input of the DWF tool is three files extracted from the Unified Bug Dataset; one contains the bug information, the other the metrics, and the third the corresponding flattened abstract syntax trees. 
In order for anyone to check our results, or to compare the solutions we present with their own, we also make these files directly available at the following link: \url{http://doi.org/10.5281/zenodo.4724941}

\section{\uppercase{Results}}
\label{sec:results}
We present the results of our experiments by giving detailed answers to the three research questions presented in Section \ref{sec:introduction}. Using these, we hope to provide insight about this representation and the use cases it might prove useful for. 

\subsection{RQ1: Is there a Doc2Vec parametrization that would produce similar or better results than learning based on code metrics?}

The first topic we investigated is the existence of a Doc2Vec parametrization for bug prediction that would generate features which are similarly expressive as using code metrics. This was very likely, since code metrics are derived from source code and thus, an adequate representation of the AST is expected to perform comparably well. To find such a representation, we tried several Doc2Vec model parametrizations and also numerous machine learning models, because so far no clearly best algorithm or direction has emerged in the literature for bug prediction.

The models we tried are usually very simple with few hyperparameter tuning possibilities, since the Doc2Vec embedding part also introduces a challenge in finding the best parameters. Combining the parameter search on the Doc2Vec embedding and on simpler machine learning models already generates a huge search space. The set of parameters for both of these tasks and the concrete setup were discussed in Section \ref{sec:experiments}.

\begin{figure}
\centering
\begin{adjustbox}{width=0.47\textwidth}
\begin{tikzpicture}
\begin{axis}
[
ytick={1,2,3,4,5,6,7,8,9,10},
yticklabels={Naive Bayes, Random Forest, SDNNC, CDNNC, Decision Tree, Linear, Logistic, KNN},
]
\addplot+[
boxplot prepared={
	median=0.26496,
	upper quartile=0.37084,
	lower quartile=0.22813,
	upper whisker=0.38664,
	lower whisker=0.20646
},
] coordinates {};
\addplot+[
boxplot prepared={
	median=0.38676,
	upper quartile=0.405715,
	lower quartile=0.33441,
	upper whisker=0.41728,
	lower whisker=0.32082
},
] coordinates {};
\addplot+[
boxplot prepared={
	median=0.42858,
	upper quartile=0.445665,
	lower quartile=0.366735,
	upper whisker=0.45085,
	lower whisker=0.34160
},
] coordinates {};
\addplot+[
boxplot prepared={
	median=0.43360,
	upper quartile=0.450275,
	lower quartile=0.41143,
	upper whisker=0.47033,
	lower whisker=0.40479
},
] coordinates {};
\addplot+[
boxplot prepared={
	median=0.34805,
	upper quartile=0.362895,
	lower quartile=0.31441,
	upper whisker=0.36766,
	lower whisker=0.30276
},
] coordinates {};
\addplot+[
boxplot prepared={
	median=0.28168,
	upper quartile=0.3092475,
	lower quartile=0.211245,
	upper whisker=0.35185,
	lower whisker=0.19349
},
] coordinates {};
\addplot+[
boxplot prepared={
	median=0.32371,
	upper quartile=0.35529,
	lower quartile=0.279615,
	upper whisker=0.40431,
	lower whisker=0.24946
},
] coordinates {};
\addplot+[
boxplot prepared={
	median=0.42028,
	upper quartile=0.44992,
	lower quartile=0.3443,
	upper whisker=0.47207,
	lower whisker=0.32502
},
] coordinates {};
\end{axis}
\end{tikzpicture}
\end{adjustbox}
\caption{Comparison of using AST embedding with various models (by F-score)}
\label{fig:emebddings_comparison}
\end{figure}
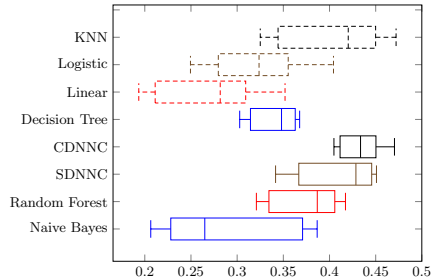

The comparative results of these algorithms can be seen in Figure~\ref{fig:emebddings_comparison}. As in every subfield of anomaly detection, accuracy as measurement would not suffice, since the dataset is highly imbalanced. Better performance metrics for such problems are recall (what proportion of the relevant class was found) and precision (what proportion of the found instances are relevant). Typically there is a trade off between the two, as was observed in this case too: some models produce precision values as high as ~0.58, but for them the recall value was around ~0.34. On the other hand, some models performed with the recall value of ~0.7, however the precision in these cases was only ~0.3. To take into account both precision and recall, we used F-score. F-score is calculated as the harmonic mean of recall and precision, therefore  it produces a value between 0 and 1. Since it's value is based on both recall and precision, and harmonic mean punishes extreme values, it gives a good intuition about a model's predictive power in imbalanced environments.  

One could argue that the best value for F-score is around 0.5, which is still low, however firstly, the available datasets are much more limited for vulnerability and bug prediction than for other fields in deep learning, which constraints the models' performance. Secondly, our main goal was to investigate the effectiveness using ast flattenings compared to metrics. Our findings regarding this question can be found on Table~\ref{table:rq1}. It can be seen that Doc2Vec vectors could indeed perform similarly, but code metrics in general still have an edge over them as their F-scores are usually higher.

\begin{table*}[t]
	\centering
	\caption{Comparison of learning based on source code embedding and metrics (values are F-scores)}
	\resizebox{0.85\textwidth}{!}{
	\begin{tabular}{ |l|c|c|c|p| }
		\hline
		Model name& Embedding & Code metrics & \begin{tabular}{@{}c@{}}Parameters \\ (vector size, window size, algorithm)\end{tabular} \\
		\hline
		Bayes \rule{0pt}{2.5ex} &  \textbf{0.414} & 0.325 & 75, 4, PV-DBOW \\
		Linear & \textbf{0.424} & 0.401 & 150, 8, PV-DBOW   \\
		Logistic & \textbf{0.425} & 0.412 &  150, 8, PV-DBOW \\
		Tree & 0.403 & \textbf{0.475} & 150, 12, PV-DBOW \\
		Random Forest & 0.441 & \textbf{0.515} & 150, 4, PV-DBOW \\
		CDNNC & 	\textbf{0.487} & 0.474 & 150, 8, PV-DBOW  \\
		SDNNC & 0.485 & \textbf{0.520} & 75, 8, PV-DM \\
		KNN & 0.491 & \textbf{0.502} & 75, 8, PV-DBOW   \\
		\hline
	\end{tabular}}
	\label{table:rq1}
\end{table*}
 
 Another observation regarding Doc2Vec parametrization is what can be found on Figure~\ref{fig:emebddings_comparison}: there is no universal best set of parameters for all of the learning models, as the distributions vary a lot in 
 their variance and in their median value, yet they are executed on the same set of Doc2Vec models. This leads us to the conclusion that each machine learning algorithm works best with a custom Doc2Vec embedding. To put it another way, different models can understand relationships based on different source code representations.

When discussing the following research questions, we build upon these conclusions.

\subsection{RQ2: Can we improve performance by combining source code embedding and code metrics?}
The next topic we discussed was the way of applying this approach. In the previous RQ we tested how source code embedding would work as a standalone method. The question that remains is how would it perform as a supplementary tool to code metric based bug prediction. Since there is no best parametrization for the source code embedding, we chose one arbitrarily from those that performed well in most of the experiments: vector size of 25, windows size of 12, learned on 80 epochs using the PV-DM algorithm. We executed all of the learning algorithms with this embedding in the following ways:

\begin{itemize}
	\item \textbf{Embedding:} Using only the chosen Doc2Vec embedding to predict bugs.
	\item \textbf{Metrics:} Using only code metrics to predict bugs.
	\item \textbf{Combined:} Combining the embedding and metrics vectors.
\end{itemize}

The results can be seen in Table~\ref{table:rq2}. The best values are denoted with bold font. As discussed earlier, even though it can be seen that in most cases the embedding by itself is close in effectiveness, code metrics still gave relatively better results, even comparable to the values found in the work of \cite{FBG20}.

\begin{table*}[t]
\centering
\caption{Comparison of different machine learning methods with the same embedding (values are F-scores)}
\begin{tabular}{ |l|c|c|c|c| }
	\hline
	Model name& Embedding & Code metrics & Combined \\
	\hline
	Bayes \rule{0pt}{2.5ex} &  0.301 & \textbf{0.325} & \textbf{0.325} \\
	Linear & 0.298 & 0.401 & \textbf{0.418}   \\
	Logistic & 0.311 & 0.412 &  \textbf{0.430} \\
	Tree & 0.374 & \textbf{0.475} & 0.461 \\
	Random Forest & 0.423 &  0.515 & \textbf{0.522} \\
	CDNNC & 	0.451 & 0.474 &\textbf{0.502}  \\
	SDNNC & 0.467 & 0.520 & \textbf{0.533} \\
	KNN & 0.463 & 0.502 & \textbf{0.524}   \\
	\hline
\end{tabular}
\label{table:rq2}
\end{table*}

However, the general pattern is that combining the two approaches, which is basically concatenating the metrics and embedding vectors is the best solution. And in cases where it is not the best (sometimes overshadowed by learning on metrics) it happens by a small margin which could probably be eliminated with a little more parameter tuning. 

Also, regarding Table~\ref{table:rq2}, it must be noted that the ``Combined'' and ``Embedding'' columns contain the values of a specific embedding (plus metrics in the case of "Combined"), which is defined earlier in this section. As discussed in RQ1, there is no best embedding for all the machine learning models, so it is possible that for some machine learning models the Embedding (and more importantly the Combined) values could be improved with further parameter search.

Based on this information we concluded that in this setup, our AST representation works best as a supplement to code metrics. Since combining the two almost always produces better or comparable results to code metrics, it is very likely that this representation provides some semantic information that the code metrics could not capture on their own, so it is indeed a useful addition for the bug prediction task.

\subsection{RQ3: Do the features provided by source code embedding give valid information or was it only random noise learned? }
The last research question that we evaluated during our experiments is the validity of our representation. On such imbalanced datasets with limited sample sizes, like in the case of bug prediction, it is always a question whether or not a given approach learns valid information. Concretely, does it capture patterns in the source that correspond to bugs, or does it only learn random noise that coincidentally works well in this case? 

We tried to answer this question by choosing a specific embedding, and supplementing it with code metrics (based on the reasoning in RQ2), then in the database we did a number of permutations on the label column. This basically randomly generates a new distribution of the classes with no semantic relation to bug prediction. Each machine learning model is then trained on these permutated datasets to see if they provide comparable results to the ones with valid bug labels. This way, each machine learning model is ``challenged'' to learn random noise from the data. 

\begin{figure}
	\centering
	\begin{adjustbox}{width=0.47\textwidth}
		\begin{tikzpicture}
		\begin{axis}
		[
		ytick={1,2,3,4,5,6,7,8,9,10},
		yticklabels={Naive Bayes, Random Forest, SDNNC, CDNNC, Decision Tree, Linear, Logistic, KNN},
		]
		\addplot+[
		boxplot prepared={
			median=0.05668,
			upper quartile=0.076495,
			lower quartile=0.035755,
			upper whisker=0.12174,
			lower whisker=0.02387
		},
		] coordinates {};
		\addplot+[
		boxplot prepared={
			median=0.073,
			upper quartile=0.097,
			lower quartile= 0.024,
			upper whisker= 0.169,
			lower whisker=0.000
		},
		] coordinates {};
		\addplot+[
		boxplot prepared={
			median=0.19246,
			upper quartile=0.199375,
			lower quartile=0.18889,
			upper whisker= 0.20545,
			lower whisker=0.17179
		},
		] coordinates {};
		\addplot+[
		boxplot prepared={
			median=0.002825,
			upper quartile=0.00463,
			lower quartile=0.00099,
			upper whisker= 0.00989,
			lower whisker=0.00049
		},
		] coordinates {};
		\addplot+[
		boxplot prepared={
			median=0.079285,
			upper quartile=0.06415,
			lower quartile=0.10100,
			upper whisker=0.075115,
			lower whisker=0.0856
		},
		] coordinates {};
		\addplot+[
		boxplot prepared={
		median=0.00121,
		upper quartile=0.00163,
		lower quartile=0.00048,
		upper whisker= 0.00241,
		lower whisker=0.00000
		},
		] coordinates {};
		\addplot+[
		boxplot prepared={
			median=0.00133,
			upper quartile=0.00169,
			lower quartile=0.00048,
			upper whisker= 0.00313,
			lower whisker=0.00024
		},
		] coordinates {};
		\addplot+[
		boxplot prepared={
			median=0.17447,
			upper quartile=0.17746,
			lower quartile=0.17261,
			upper whisker= 0.18635,
			lower whisker=0.16935
		},
		] coordinates {};
		\end{axis}
		\end{tikzpicture}
	\end{adjustbox}
	\caption{The performance of learners on datasets with permutated labels (by F-score)}
	\label{fig:permutated_comparisons}
\end{figure}
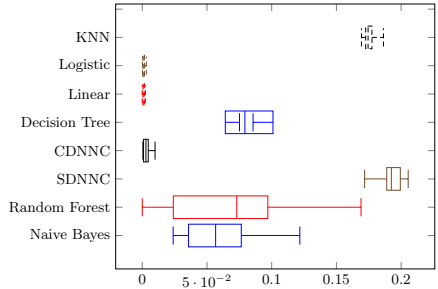

\textbf{Experimental setup:} We created 20 new databases by randomly permutating the bug label over the dataset entries. We chose to use the same embedding as we used in RQ2: Doc2Vec parameters of vector size 25, windows size of 12 on 80 epochs using DW-DM model. Our results can be seen in Figure~\ref{fig:permutated_comparisons}.

This diagram can be compared to the one in Figure \ref{fig:emebddings_comparison}, keeping in mind that in accordance with the conclusions of RQ2, the results in Figure~\ref{fig:emebddings_comparison} are nearly always worse than the combined performance with code metrics, so the difference between the distributions in Figure~\ref{fig:permutated_comparisons} and the distribution of the best performing methods (which are learning on combinations of embedding methods and code metrics) is even greater. Based on these data, it can be clearly seen that  the models did not learn well on the permutations. Even though some models could perform with minimal success, most of the results were not acceptable at all, with F-scores below 5. This leads us to the conclusion that the machine learning models generate valid results.

Also, it reinforced our reasoning about how this representation combined with code metrics indeed captures some inherent semantic information on which the downstream learning models could learn bug detection, since they could not learn random, probably non-existent relations between source codes.

\section{\uppercase{Conclusions and future work}}
\label{sec:conclusion}

In our work we presented a simple but meaningful representation of the source code based on the AST and Doc2Vec embedding that captures semantic information. We presented some background information about the literature and the efforts that have already been made in this direction, and the works that served as the backbone for our investigations. We also answered three research questions in detail, which we hope helped to understand the use of this representation and the potential in it. Based on these results we concluded that in this specific setup, using metrics and the embedding together works best when predicting bugs, and we also provided experiments on the validity of our results.

However, we also made some observations that could provide a good direction for further research:

\begin{itemize}
	\item No extensive parameter searching was done, so our results are very likely to be sub-optimal. Trying more Doc2Vec embedding and learning model parameters would probably improve the performance.
	\item Simple machine learning models were used, no real deep-learning techniques were employed on the word embedding. Considering the recent success of deep learning in NLP, using advanced architectures, for example those based on convolutional, recurrent, or attention-based neural networks could help a lot.
	\item The dataset we used is relatively small and the classes that are labeled as buggy can contain many different kinds of bugs, making effective learning very hard. Using a dataset with a bigger sample size and less variety in bug types would improve the chances of machine learning models already discussed and would facilitate the usage of deep learning models as they are typically trained on large amounts of data.
	\item Our mapping from AST to a fixed length vector using Doc2Vec relies heavily on NLP methods used on natural language texts, and thus, information could be lost by not taking more advantage of the AST's strict graph structure. A good research direction would be to use a representation method better suited for graphs, for example Graph2Vec~\cite{narayanan2017graph2vec}.
\end{itemize}

\section*{\uppercase{Acknowledgments}}
This research was partly supported by EU-funded project AssureMOSS (Grant no. 952647) and by grant 2018-1.2.1-NKP-2018-00004 ``Security Enhancing Technologies for the IoT'' funded by the Hungarian National Research, Development and Innovation Office. 

The research was also supported by the Ministry of Innovation and Technology NRDI Office within the framework of the Artificial Intelligence National Laboratory Program (MILAB) and by grant NKFIH-1279-2/2020 of the Ministry for Innovation and Technology, Hungary. 

\bibliographystyle{splncs04}
\bibliography{ast_embed}

\begin{thebibliography}{}

\bibitem[Abadi et~al., 2016]{tensorflow}
Abadi, M., Barham, P., Chen, J., Chen, Z., Davis, A., Dean, J., Devin, M.,
  Ghemawat, S., Irving, G., Isard, M., Kudlur, M., Levenberg, J., Monga, R.,
  Moore, S., Murray, D.~G., Steiner, B., Tucker, P., Vasudevan, V., Warden, P.,
  Wicke, M., Yu, Y., and Zheng, X. (2016).
\newblock Tensorflow: A system for large-scale machine learning.
\newblock In {\em 12th USENIX Symposium on Operating Systems Design and
  Implementation (OSDI 16)}, pages 265--283.

\bibitem[Analyzer, 2021]{osa}
Analyzer, O. S.~C. (2021).
\newblock \url{https://github.com/sed-inf-u-szeged/OpenStaticAnalyzer}.

\bibitem[Chen and Monperrus, 2019]{chen2019literature}
Chen, Z. and Monperrus, M. (2019).
\newblock A literature study of embeddings on source code.

\bibitem[{D'Ambros} et~al., 2010]{DLR10}
{D'Ambros}, M., {Lanza}, M., and {Robbes}, R. (2010).
\newblock An extensive comparison of bug prediction approaches.
\newblock In {\em 2010 7th IEEE Working Conference on Mining Software
  Repositories (MSR 2010)}, pages 31--41.

\bibitem[DeFreez et~al., 2018]{c_functionbased}
DeFreez, D., Thakur, A.~V., and Rubio{-}Gonz{\'{a}}lez, C. (2018).
\newblock Path-based function embedding and its application to specification
  mining.
\newblock {\em CoRR}, abs/1802.07779.

\bibitem[Devlin et~al., 2017]{python_functionbased}
Devlin, J., Uesato, J., Singh, R., and Kohli, P. (2017).
\newblock Semantic code repair using neuro-symbolic transformation networks.
\newblock {\em CoRR}, abs/1710.11054.

\bibitem[Ferenc et~al., 2020a]{FBG20}
Ferenc, R., B{\'a}n, D., Gr{\'o}sz, T., and Gyim{\'o}thy, T. (2020a).
\newblock Deep learning in static, metric-based bug prediction.
\newblock {\em Array}, 6:100021.
\newblock Open Access.

\bibitem[Ferenc et~al., 2020b]{FTL20}
Ferenc, R., T{\'o}th, Z., Lad{\'a}nyi, G., Siket, I., and Gyim{\'o}thy, T.
  (2020b).
\newblock A public unified bug dataset for java and its assessment regarding
  metrics and bug prediction.
\newblock {\em Software Quality Journal}, 28:1447–1506.
\newblock Open Access.

\bibitem[Ferenc et~al., 2020c]{FVA20}
Ferenc, R., Viszkok, T., Aladics, T., Jász, J., and Hegedűs, P. (2020c).
\newblock Deep-water framework: The swiss army knife of humans working with
  machine learning models.
\newblock {\em SoftwareX}, 12:100551.
\newblock Open Access.

\bibitem[Hammouri et~al., 2018]{HHA18}
Hammouri, A., Hammad, M., Alnabhan, M., and Alsarayrah, F. (2018).
\newblock Software bug prediction using machine learning approach.
\newblock {\em International Journal of Advanced Computer Science and
  Applications}, 9(2).

\bibitem[Harer et~al., 2018]{cpp_tokenbased}
Harer, J., Kim, L., Russell, R., Ozdemir, O., Kosta, L., Rangamani, A.,
  Hamilton, L., Centeno, G., Key, J., Ellingwood, P., McConley, M., Opper, J.,
  Chin, S., and Lazovich, T. (2018).
\newblock Automated software vulnerability detection with machine learning.

\bibitem[Jureczko and Madeyski, 2010]{JuM10}
Jureczko, M. and Madeyski, L. (2010).
\newblock Towards identifying software project clusters with regard to defect
  prediction.
\newblock In {\em Proceedings of the 6th International Conference on Predictive
  Models in Software Engineering}, PROMISE '10, New York, NY, USA. Association
  for Computing Machinery.

\bibitem[Mikolov et~al., 2013a]{mikolov2013efficient}
Mikolov, T., Chen, K., Corrado, G., and Dean, J. (2013a).
\newblock Efficient estimation of word representations in vector space.

\bibitem[Mikolov et~al., 2013b]{mikolov2013distributed}
Mikolov, T., Sutskever, I., Chen, K., Corrado, G., and Dean, J. (2013b).
\newblock Distributed representations of words and phrases and their
  compositionality.

\bibitem[Narayanan et~al., 2017]{narayanan2017graph2vec}
Narayanan, A., Chandramohan, M., Venkatesan, R., Chen, L., Liu, Y., and
  Jaiswal, S. (2017).
\newblock graph2vec: Learning distributed representations of graphs.

\bibitem[Pan et~al., 2019]{CMB19}
Pan, C., Lu, M., Xu, B., and Gao, H. (2019).
\newblock An improved cnn model for within-project software defect prediction.
\newblock {\em Applied Sciences}, 9(10).

\bibitem[Pedregosa et~al., 2011]{scikit-learn}
Pedregosa, F., Varoquaux, G., Gramfort, A., Michel, V., Thirion, B., Grisel,
  O., Blondel, M., Prettenhofer, P., Weiss, R., Dubourg, V., Vanderplas, J.,
  Passos, A., Cournapeau, D., Brucher, M., Perrot, M., and Duchesnay, E.
  (2011).
\newblock Scikit-learn: Machine learning in {P}ython.
\newblock {\em Journal of Machine Learning Research}, 12:2825--2830.

\bibitem[Puranik et~al., 2016]{PDC16}
Puranik, S., Deshpande, P., and Chandrasekaran, K. (2016).
\newblock A novel machine learning approach for bug prediction.
\newblock {\em Procedia Computer Science}, 93:924--930.
\newblock Proceedings of the 6th International Conference on Advances in
  Computing and Communications.

\bibitem[Shippey et~al., 2019]{SBH19}
Shippey, T., Bowes, D., and Hall, T. (2019).
\newblock Automatically identifying code features for software defect
  prediction: Using ast n-grams.
\newblock {\em Information and Software Technology}, 106:142--160.

\bibitem[T{\'o}th et~al., 2016]{TGF16}
T{\'o}th, Z., Gyimesi, P., and Ferenc, R. (2016).
\newblock A public bug database of {GitHub} projects and its application in bug
  prediction.
\newblock In {\em Proceedings of the 16th International Conference on
  Computational Science and Its Applications (ICCSA 2016)}, pages 625--638,
  Beijing, China. Springer International Publishing.

\bibitem[{Wang} et~al., 2016]{WLT16}
{Wang}, S., {Liu}, T., and {Tan}, L. (2016).
\newblock Automatically learning semantic features for defect prediction.
\newblock In {\em 2016 IEEE/ACM 38th International Conference on Software
  Engineering (ICSE)}, pages 297--308.

\end{thebibliography}



\end{document}